\documentclass[twocolumn,showpacs,preprintnumbers]{revtex4}
\usepackage{graphicx}
\usepackage{dcolumn}
\usepackage{bm}
\usepackage{amsmath}
\usepackage{hyperref}

\newcommand{\nabl}{\bm{\nabla}}

\begin{document}

\title{Dynamics of Skyrmion Crystal in Metallic Thin Films}
\author{Jiadong Zang$^{1,2,3}$}
\email{jdzang@fudan.edu.cn}
\author{Maxim Mostovoy$^4$}
\author{Jung Hoon Han$^5$}
\author{Naoto Nagaosa$^{2,3}$}
\email{nagaosa@ap.t.u-tokyo.ac.jp}
\affiliation{$^1$ Department of Physics, Fudan University, Shanghai 200433, China\\
$^2$ Department of Applied Physics, University of Tokyo, 7-3-1, Hongo,
Bunkyo-ku, Tokyo 113-8656, Japan\\
$^3$Cross-Correlated Materials Research Group (CMRG), and Correlated
Electron Research Group (CERG), RIKEN-ASI, Wako, Saitama 351-0198, Japan\\
$^4$Zernike Institute for Advanced Materials, University of Groningen,
Nijenborgh 4, 9747 AG Groningen, The Netherlands\\
$^5$Department of Physics, BK21 Physics Research Division, Sungkyunkwan
University, Suwon 440-746, Korea}
\date{\today }

\begin{abstract}
We study the collective dynamics of the Skyrmion crystal (SkX) in thin films
of ferromagnetic metals resulting from the nontrivial Skyrmion topology. It
is shown that the current-driven motion of the crystal reduces the
topological Hall effect and the Skyrmion trajectories bend away from the
direction of the electric current (the Skyrmion Hall effect). We find a new
dissipation mechanism in non-collinear spin textures that can lead to a much
faster spin relaxation than Gilbert damping, calculate the dispersion of
phonons in the SkX, and discuss effects of impurity pinning of Skyrmions.
\end{abstract}

\maketitle

\textit{Introduction:} Skyrmion is a topologically nontrivial
soliton solution of the nonlinear sigma model. It was noted early on
that Skyrmions in three spatial dimensions have physical properties
of baryons and the periodic Skyrmion crystal (SkX) configurations
were used to model nuclear matter\cite{skyrme1961,Klebanov_NP_1985}.
Skyrmions in two spatial dimensions play an important role in
condensed matter systems, such as quantum Hall
ferromagnets\cite{sondhi1993,Timm1998}. It was suggested that SkX
configurations can be stabilized by the Dzyaloshinskii-Moriya (DM)
interaction in ferromagnets without inversion
symmetry\cite{Bogdanov}. Such a state was recently observed in a
neutron scattering experiment in the A-phase of the ferromagnetic
metal MnSi\cite{Muehlbauer2009a}.

Recent Monte Carlo simulations indicated much greater stability of the SkX
when a bulk ferromagnet is replaced by a thin film\cite{Yi2009}. This result
was corroborated by the real-space observation of SkX in Fe$_{0.5}$Co$_{0.5}$%
Si thin film in a wide magnetic field and temperature range\cite{Yu2010a}.
Lorentz force microscopy showed that Skyrmions form a triangular lattice
with the magnetization vector antiparallel to the applied magnetic field in
the Skyrmion center and parallel at the periphery, as was also concluded
from the neutron experiment\cite{Muehlbauer2009a}.

The next important step is to explore dynamics of Skyrmion crystals
and the ways to control them in analogy to the actively studied
current- and field-driven motion of ferromagnetic domain
walls\cite{Yamaguchi2004}. Recent observation of the rotational
motion of the SkX in MnSi suggests that
Skyrmions can be manipulated by much smaller currents than domain walls\cite%
{Pf}.

In this Letter we study the coupled dynamics of spins and charges in
the SkX, focusing on effects of the nontrivial Skyrmion topology and
effective gauge fields induced by the adiabatic motion of electrons
in the SkX. We derive equation of motion for the collective
variables describing the SkX, calculate its phonon dispersion, and
discuss a new form of damping, which can be the dominant
spin-relaxation mechanism in half-metals. In addition, we consider
new transport phenomena, such as the topological Hall effect in a
sliding SkX and the Skyrmion Hall effect. We also discuss the
Skyrmion pinning by charged impurities and estimate the critical
current above which the SkX begins to slide.

\bigskip \textit{Low-energy excitations in Skyrmion crystal:} An isolated
Skyrmion has two zero modes corresponding to translations along the $x$ and $%
y$ directions. Since an applied magnetic field opens a gap in the
continuum of spin-wave excitations, the low-energy magnetic modes in
SkX are expected to be superpositions of the Skyrmion displacements,
or the phonons. The phonon modes, as well as the coupling of
Skyrmion displacements to the external current, can be consistently
described in the framework of elasticity theory.

We begin with the spin Hamiltonian $
H_{\mathrm{S}}=\int d^{3}x\left[ \frac{J}{2a}(\nabla \mathbf{n})^{2}+\frac{D%
}{a^{2}}\mathbf{n}\cdot \lbrack \nabla \times \mathbf{n}]-\frac{\mu }{a^{3}}%
\mathbf{H}\cdot \mathbf{n}\right]$, where $J$ is the exchange
constant and $D$ is the DM coupling that stabilizes the SkX
configuration $\mathbf{n}(\mathbf{x})$ in some interval of the
magnetic field $\mathbf{H}=H{\hat{z}}$\cite{Bogdanov,Han2010}. We
calculate the `harmonic
lattice energy' by considering a deformation of the SkX, $\tilde{\mathbf{n}}%
( \mathbf{x},t)=\mathbf{n}(\mathbf{x}-\mathbf{u}(\mathbf{x},t))$, where the
collective coordinate $\mathbf{u}(\mathbf{x},t)$ varies slowly at the scale
of the SkX lattice constant. The result is:
\begin{equation}\label{eq:Hamiltonian}
H_{\mathrm{S}}=d\eta J\int \frac{d^{2}x}{\xi ^{2}}[(\nabla
u_{x})^{2}+(\nabla u_{y})^{2}],
\end{equation}%
where $d$ is the film thickness and $\xi \sim a\frac{J}{D}$ is the
characteristic length scale of SkX\cite{Han2010}, with $a$ being the
lattice spacing. The dimensionless quantity $
\eta =\frac{1}{8\pi }\int_{\mathrm{uc}}d^{2}x\left( \partial _{i}\mathbf{n}%
\cdot \partial _{i}\mathbf{n}\right)$ encodes the information about
$D$ and $H$, and is called shape factor in what follows.

When an electron current is flowing through the metallic film, the
conduction electrons interact with local magnetic moments through
the Hund's rule coupling $H_{ \mathrm{H}}=-J_{\mathrm{H}}S\psi
^{\dagger }\mbox{\boldmath $\sigma$}\cdot \mathbf{n} \psi $, where
$\psi $ is the electron operator. In the case of small current
density and the Skyrmion size much larger than the Fermi wavelength
of conduction electrons, one can apply the adiabatic approximation
in which the electron spins align perfectly with the local moment.
$\psi $ is projected into the fully polarized state by $\psi =\chi
|\mathbf{n\rangle }$ with $\mbox{\boldmath $\sigma$}\cdot
\mathbf{n}|\mathbf{n\rangle =}|\mathbf{n\rangle }$. Then the
electron action $S_{\mathrm{el}}=\int dtd^{3}x[i\hbar \psi ^{\dagger
}\dot{\psi}+\frac{\hbar ^{2}}{2m}\psi ^{\dagger }\nabla ^{2}\psi
+J_{\mathrm{H}}S\psi ^{\dagger }\mbox{\boldmath $\sigma$} \cdot
\mathbf{n}\psi ]$ can be rewritten as $S_{\mathrm{el}}=\int
dtd^{3}x[i\hbar \chi
^{\dagger }\dot{\chi}-ea_{0}-\frac{1}{2m}\chi ^{\dagger }(-i\hbar \mathbf{%
\nabla }-\frac{e}{c}\mathbf{a)}^{2}\chi +J_{\mathrm{H}}S\chi^\dagger\chi]$, where $a_{\mu }\mathbf{=}%
\frac{\hbar c}{2e}(1-\cos \theta )\partial _{\mu }\varphi $ with
$\theta$ and $\varphi$ being the spherical angles describing the
direction of the local magnetization\cite{Bazaliy1998,Tatara2008}.
The gauge potential $a_\mu$  gives rise to internal electric and
magnetic fields, $\mathbf{e}$ and $\mathbf{h}$, acting on
spin-polarized electrons passing through the SkX in analogy with the
electromagnetic gauge field. Crucially, the internal magnetic field
$\mathbf{b}= \bm{\nabla}\times \mathbf{a}=\frac{\hbar c}{2e}\left(
\mathbf{n}\cdot
\partial _{x}\mathbf{n}\times \partial _{y}\mathbf{n}\right)
\hat{z}$ is intimately related to the topological charge $Q$ of
Skyrmions by\cite{Rajaraman}
$
Q=\frac{1}{4\pi }\int_{\mathrm{uc}}d^{2}x\left( \mathbf{n}\cdot \partial _{x}%
\mathbf{n}\times \partial _{y}\mathbf{n}\right) =\pm 1,
$
where the integration goes over the unit cell of the SkX. In the
language of internal gauge field, this topological feature is
nothing but the quantization of internal flux $\Omega =\int
\mathbf{h\cdot dS}$ in units of $hc/e$. The coupling of the electric
current to the internal gauge field induced by the SkX,
$H_{\mathrm{int}}=-\frac{1}{c}\int d^{3}x\mathbf{j}\cdot
\mathbf{a}$, has a simple form in terms of the collective
coordinates introduced above:
\begin{equation}
H_{\mathrm{int}}=d\frac{\hbar Q}{e}\int \!\!\frac{d^{2}x}{\xi ^{2}}
(u_{x}j_{y}-u_{y}j_{x}).  \label{eq:sd_coupling}
\end{equation}

The crucial difference between the SkX and a conventional crystal is
the form of kinetic energy. The spin dynamics originates from the
Berry phase action, $S_{\mathrm{BP}}=\frac{d}{\gamma }\int \!\!dtd^{2}x(\cos \theta -1)\dot{%
\varphi}$. Here, $\gamma =\frac{a^{3}}{\hbar (S+x/2)}$, where $x$ is
the filling of the conduction band, and $S+x/2$ is the total spin
averagely per lattice site. In terms of $\mathbf{u}$ the kinetic
energy has the form
\begin{equation}\label{eq:Berry}
S_{\mathrm{BP}}=\frac{dQ}{\gamma }\int \!\!dt\frac{d^{2}x}{\xi ^{2}}(u_{x}%
\dot{u}_{y}-u_{y}\dot{u}_{x}).
\end{equation}%
This form of the Berry phase shows that the collective variables
$u_{x}$ and $u_{y}$ describing local displacements of Skyrmions
form a pair of canonical conjugate variables, replacing $\cos \theta $ and $%
\varphi $. This characteristic property of SkX leads to several
unusual responses to applied electric currents and fields. It
originates from the Skyrmion topology and distinguishes SkX from
non-topological spin textures such as spirals and domain wall
arrays.

Using Eqs.(\ref{eq:Hamiltonian}), (\ref{eq:sd_coupling}) and
(\ref{eq:Berry}), we obtain equation of motion for $\mathbf{u}$:
\begin{equation}
\dot{\mathbf{u}}=-\frac{e\hbar \gamma }{2}\mathbf{j}+Q\frac{\gamma \eta J}{%
e\hbar }\hat{\mathbf{z}}\times \nabla ^{2}\mathbf{u}.  \label{eq:EOM}
\end{equation}%
Two consequences follow immediately. First, the dispersion of
phonons in the SkX obtained from Eq.(\ref{eq:EOM}) is quadratic,
\begin{equation}
\hbar \omega =\frac{\eta Ja^{2}}{(S+\frac{x}{2})}k^{2},
\end{equation}%
in contrast to the linear phonon dispersion in usual crystals and similar to
the dispersion of magnons in a uniform ferromagnet.
Since $u_{x}$ and $u_{y}$ play the role of the coordinate and
momentum, the longitudinal and transverse phonon modes in the SkX
merge into a single mode corresponding to the rotational motion of
Skyrmions, which leads to the quadratic dispersion. Secondly, the
SkX can move as a whole driven by the charge current $\mathbf{j}$,
with a velocity $\mathbf{V}_{\parallel
}\mathbf{=\dot{u}=}-\frac{e\hbar \gamma }{2}\mathbf{j}$. This rigid
motion of SkX leads to several interesting results discussed below.

\textit{Hall effect due to SkX motion:} In such nontrivial spin
textures, the external magnetic field (less than 0.2T for MnSi) is
more than one order of magnitude smaller than the internal one, so
that it would be neglected in what follows. As can be seen from Eq.
(\ref {eq:sd_coupling}), the collective coordinates $u_{x}$ and
$u_{y}$ play the role of electromagnetic gauge potentials $A_{y}$
and $-A_{x}$, respectively. It is thus expected that the temporal
variation of $\mathbf{u}$ induced by the current leads to a
transverse potential drop. This Hall-type effect can also be
intuitively understood using the internal magnetic field
$\mathbf{b}$ introduced above. A moving spin texture
$\mathbf{n}(\mathbf{x}-\mathbf{V}_{\parallel }t)$ induces an
internal electric field $\mathbf{e}$ analogous to the electric field
of a moving magnetic flux and related to the internal magnetic field
by $\mathbf{e}=-\frac{1}{c}\left[ \mathbf{V}_{\parallel }\times \mathbf{b}%
\right] .$ For SkX with $\mathbf{b}=b_{z}{\hat{z}}$, this electric
field
generates an electric current in the direction transverse to $\mathbf{V}%
_{\parallel }$ resulting in the Hall conductivity:
\begin{equation}
\frac{\Delta \sigma _{xy}}{\sigma_{xx} }\approx
-\frac{x}{2S+x}\frac{e\langle b_{z}\rangle \tau }{mc},
\label{eq:DynamicHall}
\end{equation}%
where $m$ is the electron mass and $\tau $ is the relaxation time. The
average internal magnetic field is $\langle b_{z}\rangle =\frac{Q\Phi _{0}}{%
2\pi \xi ^{2}}$, where $\Phi _{0}$ is the elementary flux and $2\pi
\xi ^{2}$ is the area of the unit cell of the SkX. This Hall
conductivity has the same order of magnitude as the one resulting
from the so-called topological Hall effect observed in a {\em
static} SkX\cite{Lee2009}. The latter effect is nothing but the Hall
effect induced by $\mathbf{b}$ via $\mathbf{e}=\frac{1}{c}\left[ \mathbf{v}%
\times \mathbf{b}\right] $, and $\sigma _{xy}^{Top}/\sigma
_{xx}\approx e\langle b_{z}\rangle \tau /mc$, where $\mathbf{v}$ is
the electron velocity. Our new effect differs by the factor of
$-\frac{ x}{2S+x}$ from the topological Hall effect. Its physical
origin can be easily understood by noting the total force acting on
a single conduction electron is
$\mathbf{F}=-\frac{e}{c}[(\mathbf{v}-\mathbf{V} _{\parallel })\times
\mathbf{b}]$, i.e. the Lorentz force on electrons due to the
internal magnetic field of the SkX depends on the relative velocity
of electrons and Skyrmions. When the SkX begins to slide above the
threshold electric current $j_{c}$\cite{Maekawa}, the net
topological Hall voltage will be suddenly reduced by the factor
$\frac{2S}{2S+x}$, which is how the effect of the spin-motive force
and the collective shift of Skyrmions can be identified
experimentally.

\textit{New damping mechanism and Skyrmion Hall effect:} Previously
we have systematically discussed the novel effects related to the
internal magnetic field. A natural question thus arises as to
whether there is any new
phenomena associated with the intrinsic internal electric field, which is $%
e_{i}=-\partial _{i}a_{0}-\frac{1}{c}\dot{a}_{i}=\frac{\hbar
}{2e}\left( \mathbf{n}\cdot \partial _{i}\mathbf{n}\times
\dot{\mathbf{n}}\right) $. Due to the time derivative in this
expression, its effect is absent in the static spin texture.
However, in the present case, the motion of SkX makes it
nonvanishing, and leads to an additional current $\mathbf{j}^{\prime
}$ by $\mathbf{j}^{\prime }=\sigma \mathbf{e\,}$with $\sigma$ the
conductivity of electrons. Substituting this current into the
Landau-Lifshitz-Gilbert equation\cite{Tatara2008,Bazaliy1998}
\begin{equation}
\dot{\mathbf{n}}=\frac{\hbar \gamma }{2e}[\mathbf{j}\cdot \bm{\nabla}]%
\mathbf{n}-\gamma \left[ \mathbf{n}\times \frac{\delta H_{S}}{\delta \mathbf{%
n}}\right] +\alpha \left[ \dot{\mathbf{n}}\times \mathbf{n}\right] ,
\label{eq:LLG}
\end{equation}%
the time derivative $\dot{\mathbf{n}}$ receives a correction given by%
\begin{equation}
\delta \dot{\mathbf{n}}=\frac{\hbar \gamma \sigma }{2e}(\mathbf{e}\cdot %
\bm{\nabla})\mathbf{n}=\alpha ^{\prime }\left( \mathbf{n}\cdot \partial _{i}%
\mathbf{n}\times \dot{\mathbf{n}}\right) \partial _{i}\mathbf{n}.
\label{eq:newdissipation}
\end{equation}%
The corresponding dimensionless damping constant is $
\alpha ^{\prime }=\frac{1}{(2S+x)}\frac{a^{3}\sigma }{\alpha _{\mathrm{fs}%
}\xi ^{2}c}$, where $\alpha _{\mathrm{fs}}\approx 1/137$ is the fine
structure constant. The time derivative in the r.h.s. of
Eq.(\ref{eq:newdissipation}) shows that the current induced by
internal electric field leads to dissipation. In contrast to Gilbert
damping this new mechanism does not require relativistic effects and
only involves the Hund's rule coupling that conserves the total
spin. The relaxation of the uniform magnetization, described by
Gilbert damping, is clearly impossible without the spin-orbit
coupling, which breaks the conservation of the total
spin\cite{TataraGD}. This argument, however, does not apply to
inhomogeneous magnetic textures where the breaking of the rotational
symmetry by noncollinear spin orders enables the relaxation without
the spin-orbit coupling (note that $\alpha ^{\prime }$ vanishes as
$\xi \rightarrow \infty $). Despite the non-relativistic origin,
$\alpha ^{\prime }$ depends on the DM coupling, as the latter
determines the Skyrmion size. Estimates of $\alpha ^{\prime }$ made
below show that in half-metals it can greatly exceed $\alpha $.

The effect of this new dissipation can be observed by tracing the
trajectory of Skyrmion motion. Including the new dissipation term,
the modified equation of motion (\ref{eq:EOM}) for the rigid
collective coordinates $\mathbf{u}(t)$ has the form
\begin{equation}
\dot{\mathbf{u}}=-\frac{e\hbar \gamma }{2}\mathbf{j}-Q\left( \alpha \eta
+\alpha ^{\prime }\eta ^{\prime }\right) \hat{\mathbf{z}}\times \dot{\mathbf{%
u}},  \label{eq:EOM_modified}
\end{equation}%
where the second shape factor $\eta ^{\prime }$ is given by $
\eta ^{\prime }=\frac{Q}{4\pi }\int_{\mathrm{uc}}d^{2}x\left( \mathbf{n%
}\cdot \partial _{x}\mathbf{n}\times \partial _{y}\mathbf{n}\right) \left(
\partial _{i}\mathbf{n}\cdot \partial _{i}\mathbf{n}\right)/\int_{\mathrm{%
uc}}d^{2}x\left( \partial _{i}\mathbf{n}\cdot \partial
_{i}\mathbf{n}\right)$. The new dissipation term in
Eq.(\ref{eq:EOM_modified}) is obtained by multiplying
Eq.(\ref{eq:newdissipation}) with $\partial_{j} \mathbf{n}$, using
$\dot{\mathbf{n}} = - (\dot{\mathbf{u}} \cdot \nabl) \mathbf{n}$,
and integrating over one
unit cell. The whole damping term leads to a transverse motion with velocity%
\begin{equation}
\mathbf{V}_{\perp }\approx Q(\alpha \eta +\alpha ^{\prime }\eta ^{\prime })%
\left[ \mathbf{V}_{\parallel }\times {\hat{z}}\right] .  \label{eq:SkHall}
\end{equation}%
This Skyrmion Hall effect can be observed by real-space images of
Lorentz force microscopy. The corresponding Hall angle is $\theta
=\arctan (\alpha \eta +\alpha ^{\prime }\eta ^{\prime }).$ The
estimate given below shows that main contribution to $\theta$ comes
from the new dissipation mechanism.

\textit{Pinning of Skyrmion crystal:} Next we consider the pinning
of the SkX by charged impurities. The pinning results from spatial
fluctuations of the impurity density and variations of the spin
direction in the SkX. Variations of the density of charged
impurities $\delta n_i$ give rise to local variations of the
electron density $n_e$ and since the double exchange constant $J$ is
proportional to the latter, we have $\delta J \sim J \delta
n_i/n_e$. The energy per Skyrmion $E_{\mathrm{S}} \sim J d/a$.
Denote the number of impurities in this volume by $N_1$ with
$\langle N_1 \rangle = n_i 2 \pi \xi^2 d$ and the variance $\delta
N_1 = \sqrt{N_1}$, we obtain the typical variation of the Skyrmion
energy:
\begin{equation}  \label{eq:V1}
V_1=\delta J\frac{d}{a} \sim \frac{J}{n_e 2 \pi \xi^2 a} \sqrt{N_1} = \frac{J}{n_e a \xi} \sqrt{%
\frac{n_i d}{2\pi}}.
\end{equation}
The potential energy density is then $V_0 = V_1 / (2 \pi \xi^2)$.
Substituting $n_i \sim (l a^2)^{-1}$, where $l$ is the electron mean
free path, and $n_e = \frac{x}{a^3}$, we obtain$ V_0 \sim
\frac{J}{(2\pi)^{3/2}x}\sqrt{\frac{d}{l}} \frac{a}{\xi^3}$. The
pinning regime of the whole SkX depends on the ratio of the pinning
energy $V_1$ and the elastic energy $E_{\mathrm{S}}$ of a single Skyrmion%
\cite{pin}. Let $L^2$ be the number of Skyrmions in the domain where
$u\sim \xi$. The energy gain due to the impurity pinning in the
domain is $\sim - V_1 L$, while the elastic energy cost $\sim
\frac{J d}{a}$ is
independent of the domain size. Minimizing the total energy per Skyrmion, $%
\frac{J d}{aL^2} - \frac{V_1}{L}$, we obtain $L \sim \frac{J d}{a V_1}$. $L
\gg 1$ corresponds to the case of weak (or collective) pinning of SkX, while
$L \sim 1$ corresponds to the strong pinning regime.

The pinning potential gives rise to the spin transfer torque $- Q \frac{%
\gamma \xi^2}{2d} \left[\hat{\mathbf{z}}\times \frac{\delta
V}{\delta \mathbf{u}}\right]$ in the right-hand side of
Eq.(\ref{eq:EOM}). In the steady state of moving SkX this torque has
to be compensated by the interaction with the electric current. The
critical current density is then
\begin{equation}
j_c \sim {\frac{ {e} }{\hbar}} \frac{\xi^2}{d} \biggl< {\frac{ {\partial V}
}{{\partial \mathbf{u}} }} \biggr>_{\mathrm{steady\;state}} \sim {\frac{ {e}
}{\hbar}} \frac{\xi V_0 }{d L},
\end{equation}
in the weak pinning regime, while in the strong pinning case $L$ has to be
substituted by 1. Similarly, one can estimate the gap in the spin wave
spectrum due to the pinning:
\begin{equation}
\hbar \omega_{\mathrm{pin}} \sim \frac{\hbar \gamma \xi^2}{d} \biggl< {\frac{
{\partial^2 V} }{{\partial \mathbf{u}^2} }} \biggr> \sim \frac{\hbar \gamma
\xi^2}{d} \frac{V_0L}{L^2\xi^2} = \frac{a^3}{dS}\frac{V_0}{L}.
\end{equation}

\textit{Estimates:} For estimates we consider MnSi where Mn ions
form a (distorted) cubic sublattice with $a = 2.9$ \AA. The length
of the reciprocal lattice vectors of the SkX $\sim 0.035$ \AA
$^{-1}$ corresponds to $\xi \sim 77$ \AA\ and $D \sim 0.1J$. The
kinetic energy scales as $\hbar^2/m\xi^2<J_H\sim1$eV, so that the
adiabatic approximation is justified. The electron density $n_e =
3.8 \cdot 10^{22}$ cm$^{-3}$\cite{Lee2009} corresponds to $x = n_e
a^3 \sim 0.9$ charge carriers per lattice site, while the residual
resistivity $\rho \sim 2 \mu \Omega \cdot \mbox{cm}$\cite{Lee2009}
gives an estimate of the impurity concentration $x_i \sim 5 \cdot
10^{-3}$. From the magnon dispersion in the spiral
state\cite{Grigoriev2006}, $J \sim 3$ meV.

Using these parameters we get $\alpha ^{\prime}\sim 0.1$, which
shows that the damping resulting from the electric currents
generated by non-collinear spin textures can be the dominant
mechanism of spin relaxation in half-metals, where the Gilbert
damping constant $\alpha$ is one-three orders of magnitude
smaller\cite{Kubota2009,Liu2009}. We note,
however, that since the typical electron mean free path $\xi \sim 500$ \AA\ %
is larger than the Skyrmion size $\xi$, the relation between the current and
internal electric field is nonlocal.
The topological Hall angle, $\theta_{\mathrm{H}} \sim \frac{eh_z\tau}{mc} =
\frac{1}{\alpha_{\mathrm{sf}}n_e \xi^2}\frac{\sigma}{c}$ at $T = 0$ and the
change in $\theta_{\mathrm{H}}$ induced by the sliding Skyrmion crystal [see
Eq.(\ref{eq:DynamicHall})] are also $\sim 0.1$. For a $10$nm thick film the
parameter $L \sim x \frac{\xi \sqrt{2\pi d l}}{a^2} \sim 10^3, $ i.e. the
pinning of Skyrmions by charged impurities is exceedingly weak and the
corresponding values of the pinning frequency $\hbar \omega_{\mathrm{pin}}
\sim 5 \cdot 10^{-11} \mbox{meV} $ and the critical current $j_c \sim 0.2 %
\mbox{A $\cdot$ cm$^{-2}$}$. This value is much lower than that for domain
walls\cite{Yamaguchi2004} and smaller than the ultralow threshold current $%
j_c \sim 10^{2} \mbox{A $\cdot$ cm$^{-2}$}$ observed in bulk
MnSi\cite{Pf}, which may be attributed to other pinning mechanisms
and the different dimensionality of the system. This low current
density also justifies the adiabatic approximation applied in this
work.

\textit{Comparison with vortex dynamics:} Finally, we compare the
dynamics of SkX with that of the vortex lines (VL) in type II
superconductors. A similar quadratic dispersion was obtained for VL
\cite{Fetter1967}. However, in superconductors it results from
long-ranged interactions between vortices, while the interactions
between Skyrmions are short-ranged (if one ignores the relatively
weak dipole-dipole interactions). The absence of long-range
interactions in SkX ensures the stability of the quadratic
dispersion. Furthermore, the kinetic terms in VL and SkX are
completely different. For VL $u_x$ and $u_y$ are two independent
variables, while for SkX the are conjugated variables as in
Eq.(\ref{eq:Berry}). Therefore VL are massive, while Skymions are
not. When a supercurrent flows through the type II superconductor,
the charged Cooper pairs are deflected by the VL through the Lorentz
force, which in turn gives rise to the transverse motion of the VL.
The VL dynamics is usually assumed to be
overdamped\cite{Tinkham1996}, the kinetic energy of VL is neglected,
and the Lorentz force is assumed to be counterbalanced by the
friction force. In contrast, the damping of Skyrmions is relatively
weak. The spin torque resulting from the strong Hund's rule coupling
results in a nearly longitudinal motion Skyrmion motion. The Hall
motion of VL results in a longitudinal voltage drop, which is not
important for SkX motion due to the small Hall angle.

During the completion of this paper we became aware of a recent
paper by Kim and Onoda addressed the dynamics of Skyrmions in an
itinerant double-exchange ferromagnet using a Chern-Simons-Maxwell
approach\cite{kim}. Their focus, however, seems to differ from ours.
We are grateful for the insightful discussions with Prof. Yoshinori
Tokura. This work is supported by Grant-in-Aids for Scientific
Research (No. 17105002, 19019004, 19048008, 19048015, and 21244053)
from the Ministry of Education, Culture, Sports, Science and
Technology of Japan, and also by Funding Program for World-Leading
Innovative R\&D on Science and Technology (FIRST Program). JZ is
supported by Fudan Research Program on Postgraduates. MM is
supported by the Stichting voor Fundamental Onderzoek der Materie
(FOM). HJH is supported by Mid-career Researcher Program through NRF
grant funded by the MEST Grant No. R01- 2008-000-20586-0.

\end{document}